\documentclass[aps,pre,nofootinbib,showpacs,twocolumn]{revtex4-1}%
\usepackage{amssymb,latexsym,mathrsfs}
\usepackage{amsfonts}
\usepackage{mathrsfs}
\usepackage{amsmath}
\usepackage{graphicx}
\usepackage{color}

\newcommand{\beq}{\begin{equation}}
\newcommand{\eeq}{\end{equation}}
\newcommand{\beqn}{\begin{eqnarray}}
\newcommand{\eeqn}{\end{eqnarray}}
\newcommand{\bearr}{\begin{array}}
\newcommand{\enarr}{\end{array}}

\newcommand{\ket}[1]{|#1\rangle}
\newcommand{\bra}[1]{\langle#1|}

\newcommand{\eps}{\epsilon}

\def\clr{\color{black}}
\def\bea{\begin{eqnarray}}
\def\eea{\end{eqnarray}}
\def\ba{\begin{array}}
\def\ea{\end{array}}

\def\ket{\rangle}
\def\bra{\langle}

\begin{document}
\title{Thermally driven classical Heisenberg model in one dimension }
\author{Debarshee  Bagchi}
\email[E-mail address: ]{debarshee.bagchi@saha.ac.in}
\author{P. K. Mohanty}
\email[E-mail address: ]{pk.mohanty@saha.ac.in}
\affiliation{Theoretical Condensed Matter Physics Division, Saha Institute of Nuclear Physics,
1/AF Bidhan Nagar, Kolkata 700064, India.}
\date{\today}

\begin{abstract}
We study thermal transport in a classical one-dimensional Heisenberg model
employing a \textit{discrete time odd even precessional} update scheme. 
This dynamics equilibrates a spin chain for any arbitrary temperature and finite 
value of the integration time step $\Delta t$. We rigorously show that in presence of driving
the system attains local thermal equilibrium which is a strict requirement of Fourier law. 
In the thermodynamic limit heat current for such a system obeys Fourier law for all 
temperatures, as has been recently shown [A. V. Savin, G. P. Tsironis, and X. Zotos, 
Phys. Rev. B \textbf{72}, 140402(R) (2005)]. Finite systems,
however, show an apparent ballistic transport which crosses over to a diffusive
one as the system size is increased. We provide exact results for current and energy
profiles in zero- and infinite-temperature limits.

\end{abstract}
\maketitle

\section{Introduction}
\label{Intro}
The flow of heat from a hot source to a cold sink is conventionally described in 
the hydrodynamic limit by Fourier law $J = - \kappa \, \nabla T$,
where $J$ is the steady state thermal current set up in response to the temperature 
gradient $\nabla T$ and $\kappa$ is the (finite) thermal conductivity. 
There have been several attempts \cite{Lebo,Lepri} at a microscopic `derivation' of this 
phenomenological equation which, however, could not be achieved yet.
Needless to say, in spite of the huge amount of studies over decades, 
our understanding of this basic transport phenomenon is still not quite satisfactory.
It has been found that in a variety of one-dimensional models \cite{Lebo, Lepri, AD-rev}
the thermal current scales with system size as $L^{-\alpha}$ where $\alpha \le 1$. 
This corresponds to a diverging heat conductivity in the thermodynamic
limit and thus is a violation of Fourier law. It is quite a surprise that Fourier
law, which has been remarkably consistent with experimental results in general,
is found to be invalid in many models in low dimension. 
For three-dimensional systems Fourier law is believed to be generically 
true but a rigorous proof is still lacking  \cite{AD-rev}.

In a recent work it has been shown numerically (using the Green Kubo approach \cite{Kubo})
that a classical one-dimensional Heisenberg spin model obeys Fourier law
at all temperatures \cite{savin05}. {\clr Also, it has been known for quite some
time now that at infinite temperature such spin systems follow the energy 
diffusion phenomenology in the hydrodynamic limit \cite{landau}.}
On the other hand, the 1D spin-$\frac 12$ quantum Heisenberg model (QHM)
being integrable, violates Fourier law and thermal transport is 
ballistic \cite{integr, hlubek}. For recent reviews on the theoretical 
and experimental developments in quantum spin models see \cite{QM}
and references therein.
One of the basic assumptions for the validity of Fourier law 
lies in the establishment of local thermal equilibrium (LTE) in 
the system \cite{Lebo,LTE,Lepri}, which allows one to define
thermodynamic quantities in the steady state such as pressure, temperature etc., 
locally. Most studies, however, focus on the issue of whether Fourier
law is obeyed or not, without explicitly verifying the existence of LTE. 
In fact, it is known that a system may settle down to a nonequilibrium steady state 
(NESS) which does not satisfy the essential requirement of having LTE
e.g., \textit{XY} model, Lorentz gas model \cite{AD-DD-XY}. 
It is believed that the absence of LTE in these examples is due to 
the existence of infinitely many local conserved quantities in the dynamics \cite{AD-DD-XY}.
Many of the theoretical approaches \cite{AD-rev,Lepri} rely on Linear 
response theory (Green-Kubo formula) where conductivity is 
measured by computing two point current-current time correlation
which assumes quasi-equilibrium.
The Kubo formula \cite{Kubo} is strictly valid close to equilibrium and  
in the limit $L \to \infty$, and so considerable care should be taken 
in making conclusions from experimental or simulation data which deal
with finite system size and drive \cite{AD-rev}.

In this paper, we study thermal transport properties of a classical
one-dimensional Heisenberg spin model. We use a discrete time 
odd even (DTOE) dynamics which, unlike standard numerical integration
schemes, evolves the system to the correct steady state without 
violating the required conservations. We explicitly show 
that the DTOE dynamics equilibrates a closed system and the final state is
the same for all nonzero values of $\Delta t$. With two equal temperature 
baths attached to its two ends, the system eventually equilibrates
under the DTOE dynamics and attains the temperature of the baths.
When temperature of the heat baths is different, thermal equilibrium is established in 
the system locally. With finite drive, we study in details the 
transport properties of the system e.g., thermal current $J$, energy
profiles, conductivity $\kappa$ without invoking linear response theory. 
We find that, in the thermodynamic limit, the system obeys Fourier law at all temperatures 
which is consistent with the recent study \cite{savin05}. {For a finite system however, there 
is a characteristic temperature below which the system crosses over to a
regime where transport becomes ballistic.
We present exact results for thermal current and energy profile in the limit 
$T \to 0$ and $T \to \infty$.}

The paper is organized as follows. In Sec. \ref{Model} and  Sec. \ref{DTOE} 
we describe the model and the spin dynamics in detail. We then look into the
equilibration of a closed spin chain under this dynamics in Sec. \ref{Equilibrium}. 
In Sec. \ref{Drive} we study, analytically and numerically, the transport
properties of the model in presence of thermal baths. We present a discussion and
summarize our main results in Sec. \ref{Conclusion}.

\section{Model}
\label{Model}
{Consider classical Heisenberg spins} $\{\vec{S_i}\}$ (three-dimensional unit
vectors) on a one-dimensional regular lattice of length $L$ $(1 \le i \le L)$
with periodic {boundary conditions}. The microscopic Hamiltonian is given by
\begin{equation}
\mathcal{H} = -K \sum_{i=1}^{L}\vec{ S}_i\cdot \vec{ S}_{i+1}=-K \sum_{i=1}^{L}  \cos \theta_i,
\end{equation}
where the spin-spin interaction is ferromagnetic for coupling 
$K > 0$ and anti-ferromagnetic for $K < 0$. For all the numerical results
shown in the paper $K$ has been set to unity. Here $\theta_i$ is the 
relative angle between $\vec S_i$ and $\vec S_{i+1}.$
The microscopic equation of motion can be taken as
\begin{equation}
\frac d{dt} {\vec{ S}_i} = \vec{S}_i \times \vec{B}_i,
\label{eom}
\end{equation}
where $\vec{B}_i = K(\vec{S}_{i-1} + \vec{S}_{i+1})$ is the local 
molecular field experienced by the spin at site $i$.  Clearly, 
Eq. (\ref{eom}) conserves (i) the magnitude
of the individual spin vectors $S_i^2$ and (ii) the   energy density
\begin{equation}
 {E} = \frac 1L\sum_{i=1}^{L}  \eps_i, ~~~{\rm where}~~\epsilon_i= -K \vec{ S}_i\cdot \vec{ S}_{i+1}\label{epsi}.
\end{equation}
Note that this dynamics is the classical equivalent of the quantum dynamics
for a spin-$\frac 12$ QHM. Just as the commutation relations of quantum spin
operators, the classical spins components obey the standard Poisson bracket 
relations for angular momentum.

However, there is a fundamental difference between the quantum spin-$\frac 12$
model and the classical model. The spin-$\frac 12$ QHM is integrable, whereas
all higher spin $(S \ge 1)$ QHMs (and therefore the classical model which 
corresponds to $S \to \infty$) are non-integrable. {Consequently,
there are infinitely many conserved quantities in a spin-$\frac 12$ QHM (which 
includes the energy current) and the thermal transport is ballistic \cite{integr}.
On the other hand, only the total spin and the  total  energy are conserved 
in the  corresponding classical model,  and thus one expects transport properties 
to be normal.}

Since we wish to study thermal transport, typically far away from equilibrium
(for which no general theoretical formulation is known), we need to integrate
Eq. (\ref{eom}) numerically keeping the conservations intact.
In the next section, we show that a straightforward numerical integration of 
Eq. (\ref{eom}) fails to conserve either $S_i^2$ or $E$ or both. We also
discuss in detail the advantages of using the discrete time odd even
dynamics (DTOE).

\subsection{Why DTOE dynamics?}
\label{DTOE}
\begin{figure}[htbp]
\centerline
{
\includegraphics[width=9.75cm,angle=0]{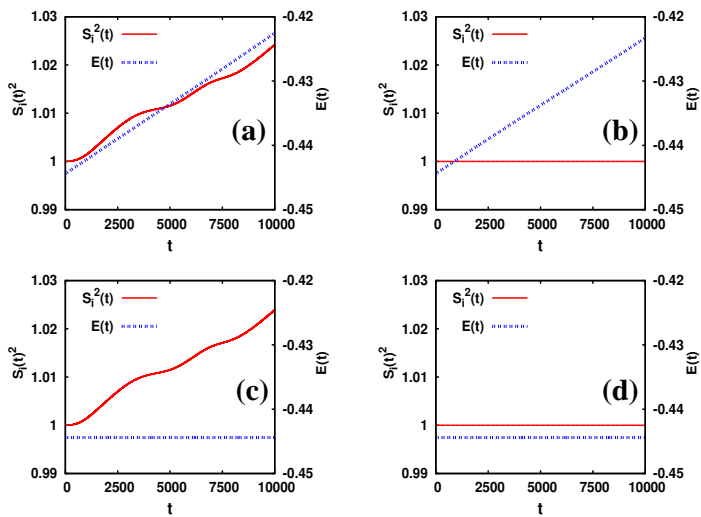}
}
\caption{(Color online) 
Evolution of spin magnitude  $S_i^2$ and the  energy density $E$.
(a) Parallel update using Eq. (\ref{f_diff}) (b) Parallel update using Eq. (\ref{precess})
(c) Odd even update using Eq. (\ref{f_diff}) (d) DTOE. For all the 
figures $\Delta t = 0.001$ and $L = 10000$. Thus, only DTOE dynamics 
conserves both $E$ and $S_i^2$.}
\label{fig:euler}
\end{figure}

To integrate the equation of motion numerically one would naively 
start off with a finite difference equation of the form
\begin{equation}
\vec{ S}_{i,t+1} = \vec{S}_{i,t} + \Delta t \,\, \left[\vec{S} \times \vec{B}\right]_{i,t}
\label{f_diff}
\end{equation}
and update \textit{all} the spins at time $t$ to obtain their values
at $t+1$. Such an \textit{Eulerian} scheme cannot be 
used for this system because of the fact that it does not satisfy the required
$S_i^2$ and $E$ conservations. It is easy to calculate the energy 
$E(t)$ and $S_i^2(t)$ in the Euler scheme using Eq. (\ref{f_diff})
which comes out to be
\begin{eqnarray}
E(t) &=& E(0) - K (\Delta t)^2 \sum_{\tau = 0}^{t-1}\sum_{i = 1}^L \left[\vec{S} \times \vec{B}\right]_{i,\tau}\cdot\left[\vec{S} \times \vec{B}\right]_{i+1,\tau} \nonumber\\ 
{S}^2_i(t) &=& S^2_i(0) + (\Delta t)^2 \sum_{\tau = 0}^{t-1}\left[\vec{S} \times \vec{B}\right]^2_{i,\tau} \nonumber\\
\label{delta_E}
\end{eqnarray}
Thus, any finite $\Delta t$, however small, breaks both the conservations and 
consequently the scheme fails for all practical purposes [see Fig. \ref{fig:euler}(a)]. 
A way to keep the magnitude of the spin vectors conserved is to use a spin 
precession dynamics

\begin{equation}
\vec{S}_{i,t+1} = \left[\vec{S} \cos \phi + (\vec{S} \times \hat{B}) \sin \phi + (\vec{S}\cdotp\hat{B})\hat{B}(1 - \cos \phi) \right]_{i,t},
\label{precess}
\end{equation}
\\
where $\hat{B}_i = \vec{B}_i/|\vec{B}_i|$ and $\phi_i = |\vec{B}_i| \Delta t$ \cite{goldstein},
instead of  Eq. (\ref{f_diff}). A spin $\vec{S}_i$, when updated using Eq. (\ref{precess}),
undergoes a precessional motion about the instantaneous local molecular field $\vec{B_i}$
which keeps its magnitude unaltered i.e., $S_{i,t+1}^2 = S_{i,t}^2$.
However, this precessional dynamics does not preserve energy conservation.
Expanding Eq. (\ref{precess}) in powers of $\Delta t$ and retaining terms up to 
$\mathcal{O}(\Delta t)$ one obtains back the first equation of Eq. (\ref{delta_E})
and thus energy conservation still remains violated for any $\Delta t > 0$.
This has also been shown numerically in Fig. \ref{fig:euler}(b). Other numerical schemes 
such as Runge-Kutta, predictor corrector method etc., will also fail to preserve the energy
conservation for the same reason.

We now describe an odd-even update rule which along with 
the precessional dynamics has been herein referred to as the DTOE dynamics.
Starting from a spin configuration $\{\vec{S}_i\}$, we numerically
implement the dynamics described in Eq. (1) by alternate parallel
updates of the spins on odd and even sublattices.
Henceforth, we refer to these two groups of spins as 
\textit{odd} and \textit{even} spins. At each Monte 
Carlo step (MCS), first only even spins are updated using the 
spin precession dynamics Eq. (\ref{precess}) and the odd spins 
are kept unaltered. Next, the spins on the odd sublattice are updated
similarly. These two steps update all the spins $\{\vec S_i\}$ in the 
system and constitute one MCS.
It is straightforward to check that update of any spin 
$\vec{S}_i$ affects only the energy of the neighboring bonds 
$\eps_{i-1}$ and $\eps_i$ but their sum ($\eps_{i-1} + \eps_i$)
remains constant. Thus DTOE dynamics is strictly energy conserving.

Clearly, the spin precession dynamics conserves the magnitude of the spin 
vectors while, energy conservation is maintained by the odd-even update
rule (see Fig. \ref{fig:euler}). A recent paper \cite{DTOE} has also
employed this odd-even precessional dynamics with large $\Delta t$ to study
transport in a classical Heisenberg model (1D periodic spin system)
in presence of quenched disorder numerically. 
Although this dynamics does not directly follow from the equation of 
motion [Eq. (\ref{eom})], it can be used to study the system numerically
provided that the system equilibrates for any arbitrary $\Delta t$.
In the next section, we study the equilibration of a closed system when
evolved using DTOE dynamics.

\section{Closed system}
\label{Equilibrium}
We first investigate whether a closed system (i.e., with periodic boundary
conditions) under DTOE dynamics evolves to the correct steady state for 
different values of $\Delta t$. To do this, first we compute the correlation
functions of the system in canonical ensemble, subjected to temperature $T$ and 
then show numerically that the same correlation functions are obtained from a 
closed system with a fixed energy (i.e. in a micro-canonical ensemble). 

The partition function of the  system \cite{Joyce} with the Hamiltonian
given by Eq. (\ref{eom}) is
\begin{equation}
\mathcal{Z} = \int \prod_{i=1}^L \left(\frac{d\vec{S}_i}{4\pi}\right) \exp(\beta K \sum_{i=1}^L \vec{ S}_i\cdot \vec{ S}_{i+1} ),
\end{equation} 
{$\beta = 1/k_BT$ and $k_B$ has been set equal to unity henceforth. }
The two-spin correlation functions are therefore given by

\begin{equation}
 C_{lr} = \langle P_l(\vec{S}_i \cdotp \vec{S}_{i+r})\rangle = \left[ \frac{\lambda_l(\beta K)}{\lambda_0 (\beta K) }\right]^r,
\end{equation}

where $P_l$ are Legendre polynomials and

\begin{equation}
 \lambda_l (k) = \frac 12 \int_{-1}^1 P_l(x) e^{kx} dx.
\end{equation}
These correlation functions can be written explicitly in terms of the Langevin function 
${\mathcal L}(x)$,  for example,

\begin{equation}
C_{1r} = \left[{\mathcal L}(\beta K)\right]^r \;\;\;\;\;\;\;\;\;
C_{2r} = \left[1 - 3 \mathcal{L}(\beta K)/(\beta K)\right]^r.
\label{Clr}
\end{equation}
It is evident that average energy of the system is $ - K C_{11}$ and thus, a closed 
system with a fixed energy density $E$ has an effective $\beta = {\mathcal L}^{-1} (-E/K)/K$.

\begin{figure}[htbp]
\centerline
{
\includegraphics[width=9.25cm,height=9.5cm,angle=-90]{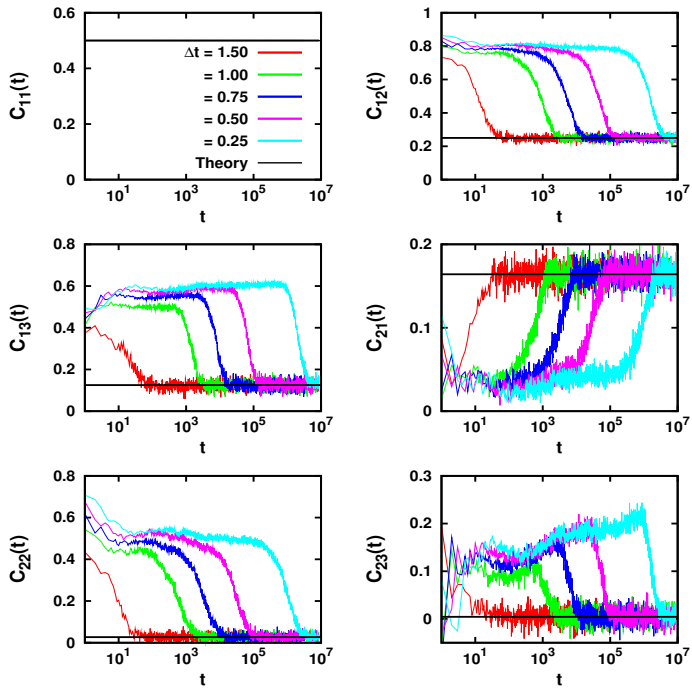}
}
\caption{(Color online) Semi-log plots for the evolution of equal-time correlations 
$C_{11}$, $C_{12}$, $C_{13}$, $C_{21}$, $C_{22}$, $C_{23}$ as obtained from 
simulation for a closed system of $L = 1000$. DTOE dynamics has been used
with time-step $\Delta t = 1.5,1.0,0.75,0.5, 0.25$ and $E = -0.5$.
}
\label{fig:mce_ttl}
\end{figure}

\begin{figure}[htbp]
\centerline
{
\includegraphics[width=6.50cm,angle=-90]{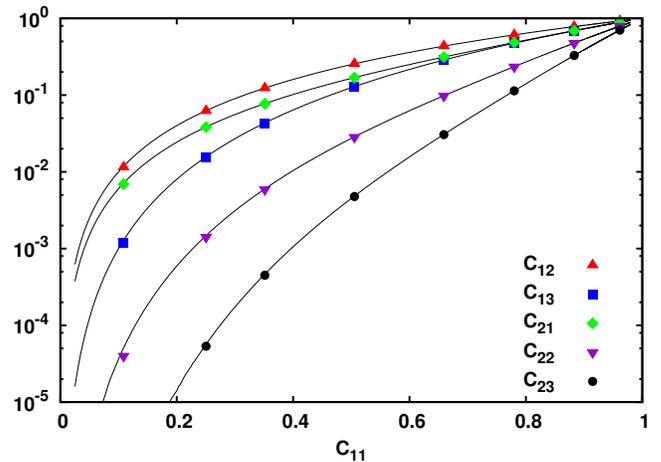}
}
\caption{(Color online) Semi-log plot for different equilibrium correlations 
$C_{lr}$ for $l = 1, 2$ vs. $ C_{11} = -E$
obtained from simulation (points) and compared with theory (lines) for a 
periodic lattice of $L = 1000$. DTOE dynamics has been used with time-step $\Delta t $ = 1.0.}
\label{fig:corr_mce}
\end{figure}

First, let us compute the equal-time spin-spin correlations
$C_{lr}(t)$ for a closed system and check that these evolve to 
the stationary value given by Eq. (\ref{Clr}).
The time series for $C_{1r}(t)$ and $C_{2r}(t)$ with different 
$r$ values are shown in the Fig. \ref{fig:mce_ttl}. We find that
the correlation functions for different $\Delta t$ saturate to
the same value at late times.
In Fig. \ref{fig:corr_mce} the time averaged equilibrium correlation 
functions $C_{lr}$ obtained from systems set at different energies,
are also found to be in remarkable agreement with Eq. (\ref{Clr}).
Thus the DTOE dynamics equilibrates the system irrespective of the value of
$\Delta t$; $\Delta t$ only alters the equilibration time of the system.
A larger $\Delta t$ is preferable as equilibration in this case is 
attained faster.

\section{Open system}
\label{Drive}
  
\subsection{Modeling heat bath}
In order to study energy transport in the system, we now look into an
open system with heat baths attached to its two ends. 
The left and right baths are set at temperatures $1/\beta_l$
and $1/\beta_r$ respectively. Each bath is known as a 
\textit{stochastic thermal bath} \cite{bath} which means that 
it is in equilibrium at its respective temperature and has a 
Boltzmann energy distribution. The baths are implemented by 
introducing two additional sites $i = 0$ and $i = L+1$ in the
system with spins $\vec{S}_0$ and $\vec{S}_{L+1}$ respectively. 
These pairs of spins ($\vec{S}_0, \vec{S}_1$) and 
($\vec{S}_L, \vec{S}_{L+1}$) behave as stochastic heat
baths at two ends of the system.
The baths are in equilibrium at their respective temperatures 
and the bond energies $\eps_0$ and $\eps_L$ have a Boltzmann
distribution
\begin{equation}
P(\eps_0) \sim e^{-\beta_l \epsilon_0}  ~ {\rm and} ~  P(\eps_L) \sim e^{-\beta_r \epsilon_{_L}}.
\label{dist}
\end{equation}
The interaction strength of the bath spins with the system is taken to 
be $K$, and therefore both $\eps_0$ and $\eps_L$ are bounded in the range $(-K, K)$.
Thus the mean energies of the left and the right bath are given by
$$E_{l} =  \langle \eps_{0} \rangle  = - K \mathcal{L}(\beta_{l} K) ~ {\rm and}~ E_{r} 
=  \langle \eps_{_L} \rangle  = - K \mathcal{L}(\beta_{r} K).$$

Following the odd-even rule, the spin $\vec{S}_0$ is updated along with 
the even spins, whereas $\vec S_{L+1}$ is updated with odd (even) spins 
depending on whether $L$ is  even (odd).
To update $\vec{S}_0$, first the energy of the bond $\epsilon_0$ between the spins 
$(\vec{S}_0,\vec{S}_1)$ is set to a value drawn randomly from  $P(\eps_0)$ 
given in Eq. (\ref{dist}). 
The spin $\vec{S}_0$ is then constructed such that $\eps_0 = -K\vec{S}_0 \cdotp \vec{S}_1.$
During this update $\vec{S}_1$ is not modified as it belongs to the odd sublattice.
At the right end, the spin $\vec{S}_{L+1}$ is updated similarly. We must mention 
that energy conservation is violated during update of bath spins. 
The interaction of the bath spins with the neighboring spins allow
boundary fluctuations to propagate into the bulk, thus inducing a
thermal current in the system.
 
\begin{figure}[htbp]
{
\includegraphics[width=4.25cm,angle=0]{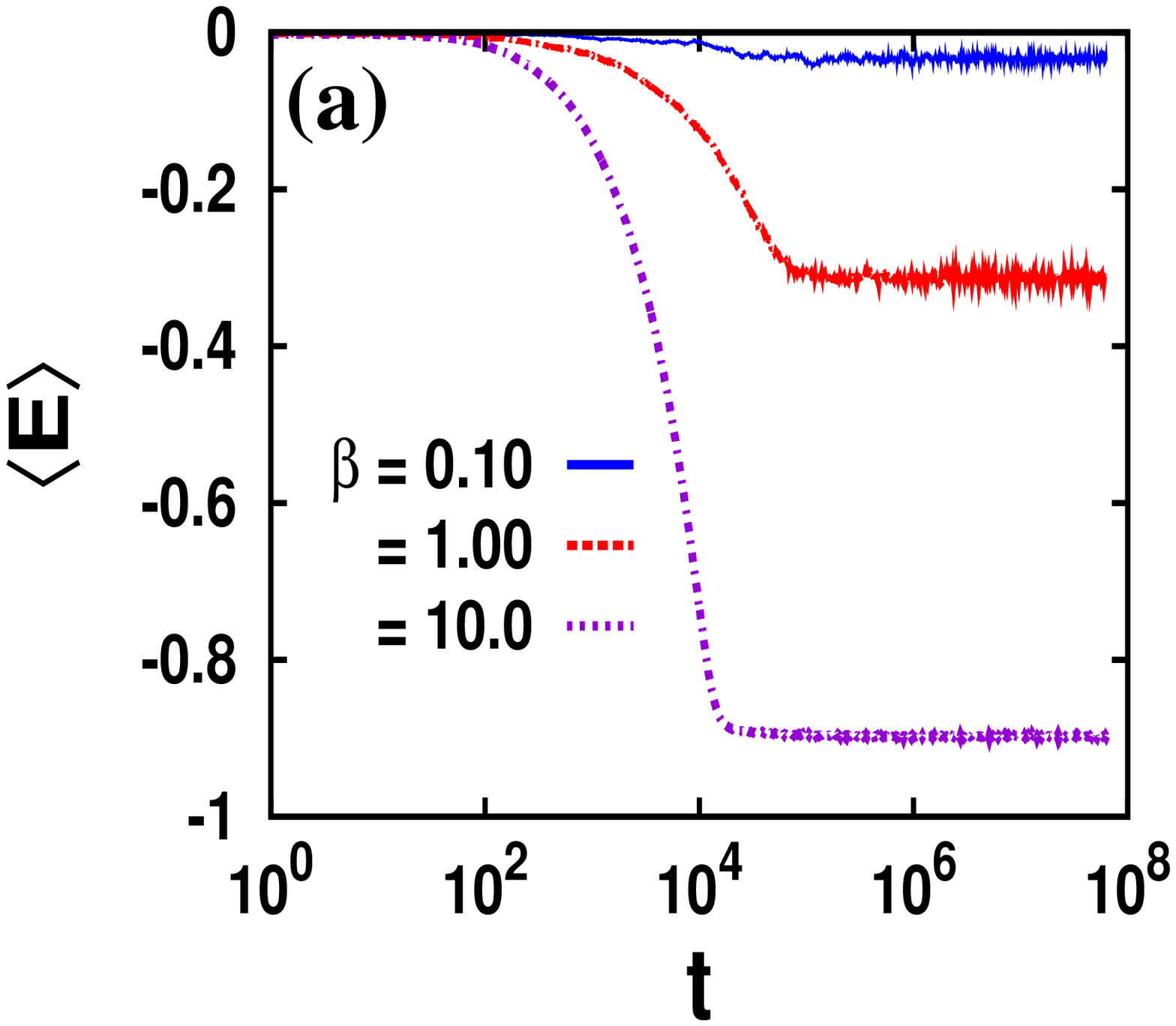}
\includegraphics[width=4.25cm,angle=0]{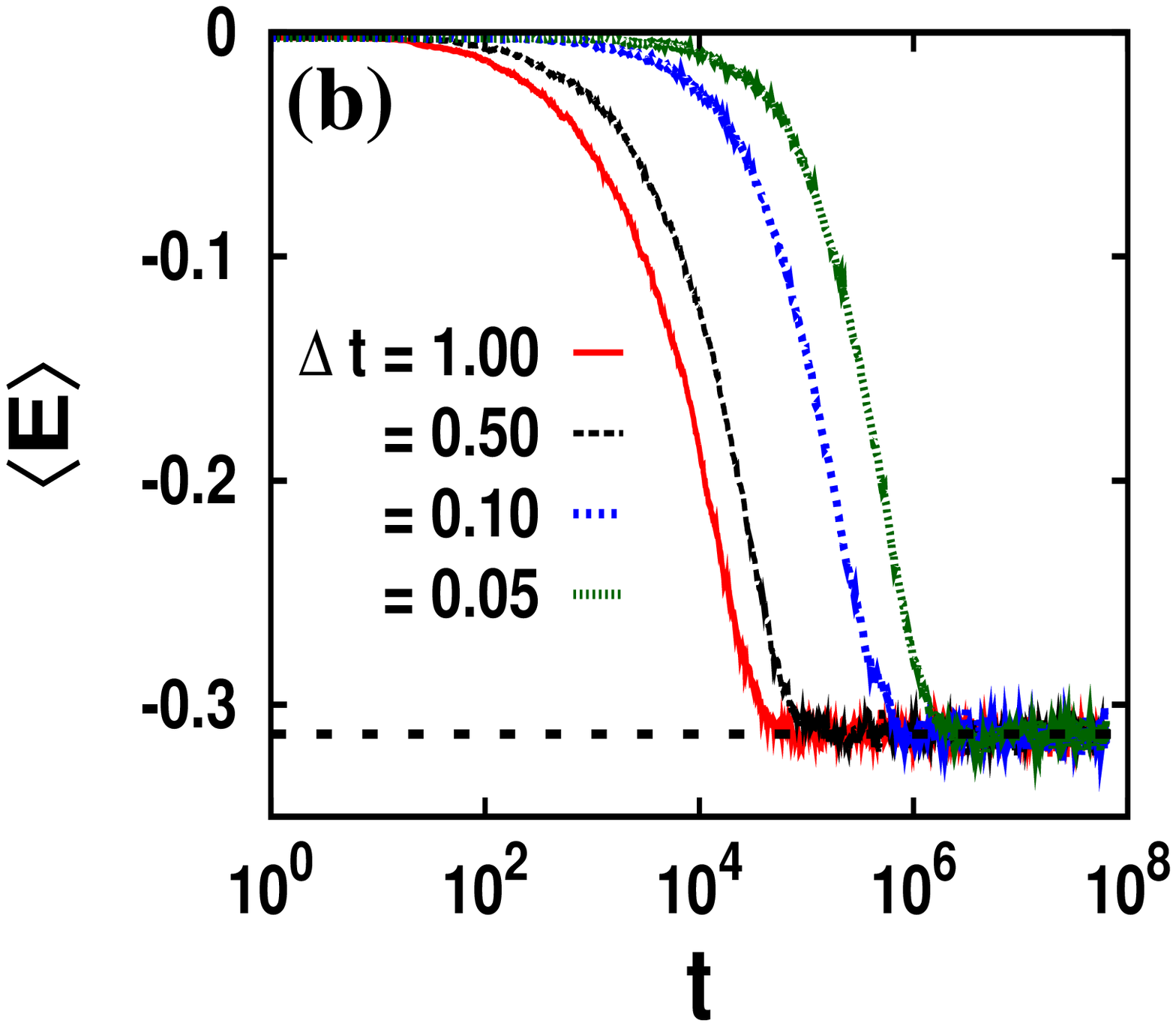}
\includegraphics[width=8.7cm,angle=0]{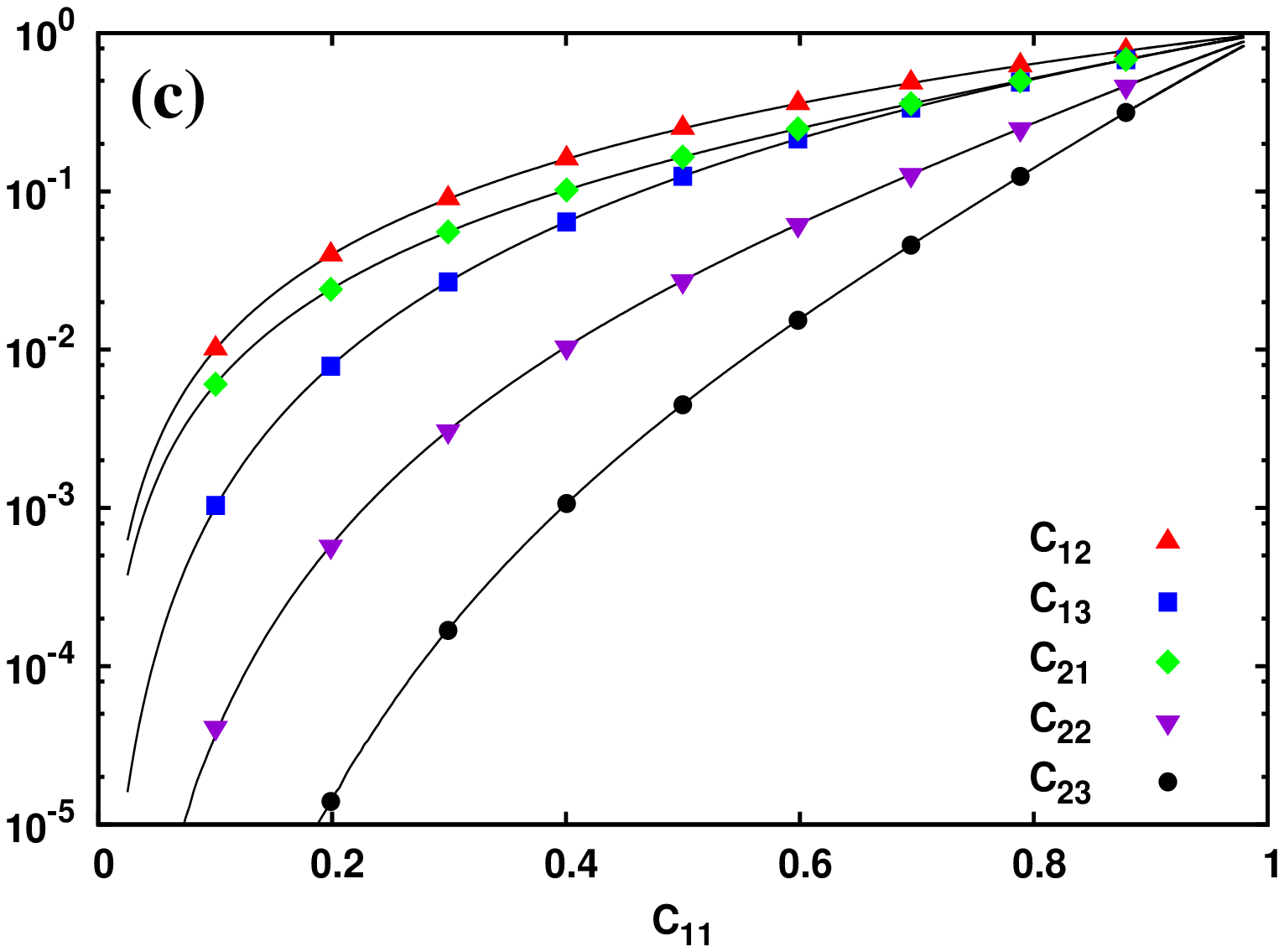}
}
\caption{(Color online) Semi-log plot for the evolution
of average energy $\langle E \rangle$ for a system with
$\beta_l=\beta_r=\beta$ for
(a) $\beta = 0.1,1.0,10.0$ and a fixed  $\Delta t = 1.0$ and 
(b) $\Delta t = 1.0, 0.5, 0.1, 0.05$ and a fixed $\beta=1.0.$
(c) Different correlations $C_{lr}$ obtained from simulation 
(points) are shown as a function of $C_{11}$ along with the
functions Eq. (\ref{Clr}) (lines); $L = 1000$ and $\Delta t  = 1.0.$
}
\label{fig:corr_ce}
\end{figure}

Although the closed system equilibrates under DTOE dynamics, it is not
guaranteed that an open system will also equilibrate when baths are 
attached to its two ends. 
Before studying the system with a finite drive, we study the equilibration
of an open system with baths maintained at the same temperature, thus 
still keeping the system in equilibrium at temperature $T$, i.e. 
$\beta_l=\beta=\beta_r.$ With baths maintained at equal temperature $T$,
the spin chain is expected to eventually reach a thermodynamic equilibrium
corresponding to the bath temperature $T$.  
We have calculated numerically the average energy of the system  
$\langle E \rangle$ for different values of $\beta= 10, 1,$ and $ 0.1$, 
which is shown in Fig. \ref{fig:corr_ce}(a).
Evidently, at late times $\langle E \rangle$ approaches the stationary value 
$-K{\cal L} (K\beta)$.
Figure \ref{fig:corr_ce}(b) shows that the system attains a unique stationary
state consistent with the bath temperature and this final state is independent
of the value of $\Delta t$ used. Similar to the case of the closed system,
$\Delta t$ decides only the equilibration time of the system.
We also measure the  equilibrium correlations $C_{lr}$ which are shown in 
Fig. \ref{fig:corr_ce}(c) where the numerical values are found to be 
in agreement with Eq. (\ref{Clr}).
This assures us that for any nonzero $\Delta t$ the DTOE dynamics is
no different from the equation of motion (\ref{eom}), and allows the
system to attain the correct equilibrium state.

\subsection{LTE in a driven system}
A finite thermal drive is imposed on the spin chain by setting the two heat baths
at unequal temperature, i.e. $\beta_l\neq \beta_r$. The bath and bulk spins are 
updated as mentioned in the previous section.
Now, since the bath temperatures are unequal, the system is driven out of 
equilibrium.
However, it may still be possible to define a temperature like thermodynamic variable
\textit{locally} in a region if the average energy of the sites $\langle \eps_i \rangle$
belonging to that region is not too different from each other. 
The system is said to have local thermal equilibrium (LTE) if all the correlation
functions measured in this local region are identical to those of a thermodynamically
large equilibrium system with an average energy $\langle \eps_i \rangle$.

\begin{figure}[htb]
\centerline
{
\includegraphics[width=6.50cm,angle=-90]{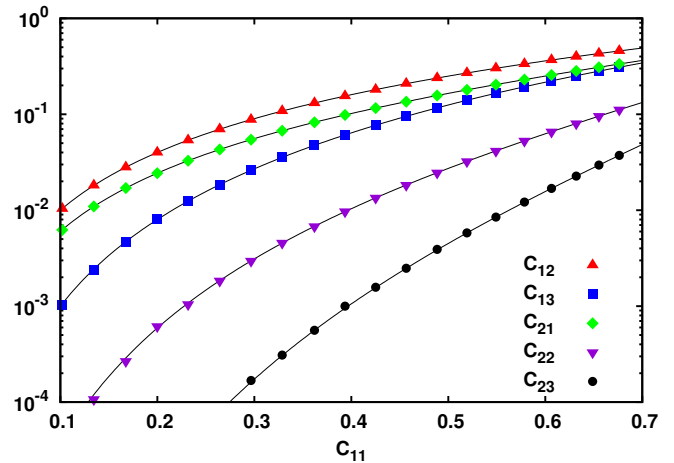}
}
\caption{(Color online) Semi-log plot for different correlations $C_{lr}(x)$,
where $0 < x = i/L < 1$, for $l = 1, 2$ vs. $ C_{11}(x)$ obtained from 
simulation (points) and compared with Eq. (\ref{Clr}) (lines) for a 
open system of $L = 1000$. The average bath energies are $E_l = -0.7$ and
$E_r = -0.1$. DTOE dynamics has been used with time-step $\Delta t = 5.0$.
}
\label{fig:corr_neq}
\end{figure}

Numerically, one measures the correlation functions $C_{lr}(x)$ locally 
over $n \ll L$ consecutive sites about $x = i/L$ such that the average 
energy of these $n$ sites is almost equal to each other.
For a system of size $L = 1000$ we measure $C_{lr}(x)$ up to three nearest neighbors
for $l = 1, 2$ and by averaging them over $n=20$ sites.
This is shown in Fig. \ref{fig:corr_neq}, where we have shown $C_{lr}(x)$ 
for different $x$ (in the range $(0,1)$) as a parametric function of $C_{11}(x)$.
For comparison, the equilibrium curves (from Eq. (\ref{Clr})) are also shown 
in the figure as solid lines. 
An excellent match with the equilibrium functions assures that the driven
spin system attains thermal equilibrium locally.
Thus, following the equilibrium definition, we may define {\it uniquely} 
the local inverse temperature

\begin{equation}
\beta(x) \equiv \frac 1 K {\cal L} ^{-1} (C_{11}(x)).
\label{betax}
\end{equation}

\subsection{Analytical Results}
In the previous section we have seen that the driven system attains 
local thermal equilibrium, and thus a local temperature can be uniquely defined
through Eq. (\ref{betax}).
Therefore, the usual definition of Fourier law $J \propto \nabla T$ 
can also be equivalently expressed as $ J \propto \nabla E.$
The thermal current and the energy profile are measured as described in the following.
Since the DTOE dynamics alternately updates only half of the spins 
(but all the bond energies simultaneously),
the energy of the bonds $\epsilon^o_i$ measured immediately after 
the update of odd spins is different from $\epsilon^e_i$ measured 
after the update of even spins.
Clearly, this difference $\epsilon^e_i - \epsilon^o_i$ is a measure 
of the energy flowing through the $i$-th bond in each MCS.
Thus the thermal current in the steady state is given by
\begin{equation}
J = \langle \epsilon^e_i - \epsilon^o_i  \rangle
\label{J}
\end{equation}
and the average energy of $i$-th bond is 
$\epsilon_i = \frac 12 \langle \epsilon^e_i + \epsilon^o_i \rangle$.
{\clr In fact, this expression for the current, in the limit $\Delta t \to 0$, is consistent
with that obtained from the continuity equation $\dot \epsilon_i(t) = J_{i-1}(t) - J_i(t)$
where $\epsilon_i(t) = - K \vec{ S}_i\cdot \vec{ S}_{i+1}$
is the local energy density. Straightforward
calculation using the continuity equation gives
the instantaneous current across $i$-th bond
\begin{equation}
J_i(t) = K ~ \vec S_i \cdot (\vec S_{i+1} \times \vec S_{i+2}).
\label{contEq}
\end{equation}

Again, when the $i$-th site (say, even) gets updated, the energy of the $i$-th bond 
is $\epsilon_i^e = - K ~\vec S_i(t+1) \cdot \vec S_{i+1}(t)$ and after 
the subsequent update of odd sites  it   becomes  $\epsilon_i^o = - K ~\vec S_i(t+1) \cdot \vec  S_{i+1}(t+1)$. 
Thus using Eq.  (\ref{precess}), the instantaneous current (time is measured
in units of $\Delta t$) across the $i$-th bond in  the limit  $\Delta t \to 0,$
reduces to 
\begin{equation}
 J_i(t) =  \epsilon^e_i - \epsilon^o_i = K ~ \vec S_i \cdot (\vec S_{i+1} \times \vec S_{i+2}),
\end{equation}
which is  same as Eq. (\ref{contEq}).}

In the following, we show that the model with DTOE dynamics can be solved exactly
in both $T \to 0$ and $T \to \infty$ limits to obtain 
analytical expressions for the energy current and energy profile.
We show that energy transport in any finite system 
is ballistic in the limit $T \to 0$, whereas diffusive
transport is observed for $T \to \infty.$
%
%
%
\vskip0.5cm
{\it Ballistic Limit} ($T\to 0$):
In this limit, the spins are nearly aligned (hence  $\vec{S}_i \times \vec{B}_i \simeq 0$)
and, therefore, precess by small angles. In DTOE dynamics, the angle of precession is
$\phi_i = |\vec{B}_i| \Delta t$ and so for small $\Delta t$ it mimics the
dynamics of the system at low temperature. This equivalence can be utilised to write 
%
%
%
%
%
%
%
%
%
%
%
the energy function Eq. (\ref{eom}) in this limit as
\begin{equation}
\mathcal{H} \simeq - K \sum  (1- \frac 1 2 \theta_i^2),
\end{equation}
which is similar to the energy function of a harmonic system.
Since the DTOE dynamics is energy conserving, update of a spin $\vec S_i$ assures that 
$\theta_{i-1}^2 + \theta_i^2$ remains invariant. Again, since the low temperature stationary 
state dynamics is governed by spin waves i.e., spin configuration whose orientation
varies slowly with distance along the axis of the chain, the spins are locally parallel
and $\theta_{i-1}  + \theta_i$ is invariant in the $\Delta t \to 0$ limit. 
Thus the only allowed dynamics for the angle variable is
\begin{equation}
\theta_{i-1, t}  \to  \theta_{i,t+\Delta t}  ~~~~~ \theta_{i,t}  \to  \theta_{i-1,t+\Delta t}. 
\end{equation}
Consequently, the energy of the bonds that connect to the $i$-th spin, namely 
$\epsilon_{i-1}$ and $\epsilon_i$, are mutually exchanged in this limit. 
Starting from $t = 0$, the bond energies $\epsilon_{2i+1}$ for the odd 
bonds `move' to the right and the even bond energies $\epsilon_{2i}$ `move' 
to the left ballistically (without any scattering).

In the steady state, the average bond energies after the update of odd spins
are $\langle \epsilon_{2i}^o\rangle = E_l$ and $\langle\epsilon_{2i+1}^o\rangle = E_r$
whereas, the same after the update of even spins become 
$\langle \epsilon_{2i}^e\rangle = E_r$ and $\langle\epsilon_{2i+1}^e\rangle = E_l$.
Thus current $J =  E_l - E_r$ is independent of the system size $L$ and 
thermal transport is ballistic. In the steady state, the temperature is same
at all bulk sites, which is given by
\begin{equation}
T_{bulk} = \frac K {\mathcal{L}^{-1}\left(-\frac {E_l + E_r}{2K}\right)}
\end{equation}
Note that this ballistic behavior is a consequence of the fact that
the limit $\Delta t \to 0$ is taken before the thermodynamic limit 
$L \to \infty$. When $\Delta t \simeq 0$, the spins precess 
slowly since the precision angle $\phi_i$ is proportional to
$K \Delta t$. Thus the effective correlation length diverges as $\Delta t \to 0$ and 
energy in any finite system would be transferred to arbitrary 
distances without being scattered.

Also, unlike the equilibrium case, the correlation functions
of the  driven system \textit{depend} on $\Delta t$ when $L$ is finite.
In the following, we argue that
when $\Delta t\to 0$ the correlation length $\xi$ actually diverges. 
Since the energy profile in this limit is flat, a small change 
$\Delta t \to \Delta t'$ will not change the correlation functions substantially. 
In order to keep the correlation functions (which are functions $\beta K$) unaltered, 
the inverse temperature $\beta$ should scale as $\beta' = \beta \Delta t/\Delta t'$,
so that $K' \beta'  = K \beta.$ Again, since the correlation length
$\xi = \frac{1}{|\ln {\cal L}(K \beta )|}$ (calculated from Eq. (\ref{Clr}) taking
$C_{1r} \equiv e^{-r/\xi}$) diverges linearly with $\beta$
in the limit $\beta \to \infty$, we have
\begin{equation}
\xi \sim (\Delta t)^{-1}.
\label{xiDt}
\end{equation}
{\clr
This indicates that the steady state energy profile also depends on
$\Delta t$, which will be discussed later in section \ref{Drive} E (see Fig. \ref{fig:prof_ELdt}(a)).
We must mention here that this $\Delta t$ dependence is only a numerical artifact
in finite systems. In fact, $1/\Delta t$ has to be compared with the two other
length scales of the problem, namely, the size of the system $L$ and $1/T$
(as the correlation length also diverges in $T \to 0$ limit) 
and thus, the effective correlation length will appear to be $1/\Delta t$
only when both $L$ and $1/T$ are much smaller. In other words, for the numerical
integration of Eq. (\ref{eom}), one must choose the integration time step
$\Delta t$ larger than both $1/L$ and $T$ to avoid dependence of the steady
state on $\Delta t.$ Therefore, a thermodynamically large system in
this problem corresponds to a system with $L \gg 1/T \gg 1/\Delta t.$
In this limit, the correlation length $\xi$ remains smaller than
$L$ for all $T>0$; the steady state behaviour is independent
of $\Delta t$ and one recovers diffusive thermal transport
(see Fig. \ref{fig:m} and related discussions later).}
\vskip0.5cm
{\it Diffusive limit} ($T\to \infty$):
%
%
In the other limit $T\to \infty$,  the spin orientation is random and thus the dynamics is 
equivalent to large $\Delta t$ limit, where the precession angle $\phi$ is large
and effectively the spin precesses by a random angle.
Updating the $i$-th spin then results in random re-sharing of
the bond energies $\epsilon_{i-1}$ and $\epsilon_i$ obeying the local energy conservation
imposed by the DTOE dynamics. Effectively, 
\bea
\epsilon_{i-1,t+1}= r (\epsilon_{i-1} + \epsilon_i)_t  &~~~~&
\epsilon_{i,t+1}= (1-r) (\epsilon_{i-1} + \epsilon_i)_t \nonumber\\
\eea
where, $r$ is a uniform random number in the range $(0,1)$. This dynamics is similar to
the diffusive dynamics discussed by Kipnis \textit{et. al.} \cite{kipnis} except the
fact that here we use the DTOE dynamics. 
In the  steady state, the average energies  at different sites satisfy the following equations. 
Update of odd sites ensures that for $j=0,1,\dots,L/2,$ 
\bea
\bra \epsilon_{2i}^o \ket &=&  \left( \bra \epsilon_{2i}^e\ket  + \bra\epsilon_{2i+1}^e\ket\right)/2  \cr
\bra\epsilon_{2i+1}^o\ket &=&  \left(\bra \epsilon_{2i}^e\ket +\bra\epsilon_{2i+1}^e\ket\right)/2 
\label{odd}
\eea
Similarly update of odd sites gives 
\bea
\bra\epsilon_{2i-1}^e\ket &=&  \left( \bra\epsilon_{2i-1}^o\ket  + \bra\epsilon_{2i}^o\ket\right)/2  \cr
\bra\epsilon_{2i}^e\ket &=&  \left( \bra\epsilon_{2i-1}^o\ket +\bra\epsilon_{2i}^o\ket\right)/2, 
\label{evn}
\eea
for  $j=1,2,\dots,L/2$, along with the  boundary conditions
\begin{equation} 
\bra\epsilon_{0}^e\ket = E_l  ~~~~  \bra\epsilon_{L+1}^e\ket = E_r.\label{bc}
\end{equation} 
These set of linear equations (\ref{odd})-(\ref{bc}) provide a unique solution 
\bea
\bra\epsilon_{ i}^o\ket &=&  E_l +  \frac{E_r  - E_l}{L+2} \left(i+ 
\eta \right) \cr
\bra\epsilon_{ i}^e\ket &=&  E_l +  \frac{E_r  - E_l}{L+2} \left( i+ 1-\eta \right),
\eea
where $\eta  =0,1$ for   $i=$ even, odd respectively.
Clearly the energy profile $\epsilon_i = (\bra \epsilon_i^o +  \epsilon_i^e\ket) /2$ is  
linear and the current $J = \frac{E_r  - E_l}{L+2}$ follows  Fourier law.

\subsection{Thermal current}
\begin{figure}[htbp]
\centerline
{
\includegraphics[width=6.5cm,angle=-90]{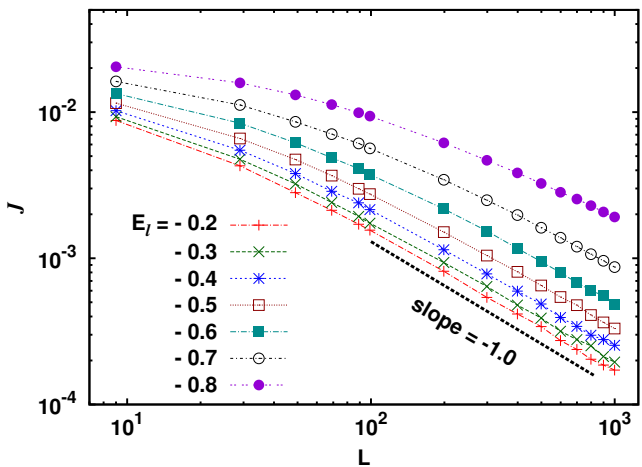}
}
\caption{(Color online) Log-log plot of the steady state current $J$ vs. $L$ 
agrees well with Fourier law $J \sim L^{-1}$ for large $L$, but deviates for 
small $L$ values. The average energy of the two baths are $E_l$ and 
$E_r = E_l + \Delta E$ with $\Delta E  = 0.1$ and $\Delta t = 1.0$.
}
\label{fig:JLlogsc}
\end{figure}

For finite $T$, the model is not analytically solvable and we study transport 
properties numerically using DTOE dynamics. Two thermal baths are
attached to the two ends of the system having average energy $E_l = E$
and $E_r = E + \Delta E$ respectively.
The steady state current $J$, measured using Eq. (\ref{J}), is shown in  
Fig \ref{fig:JLlogsc}. Clearly, $J$ decreases with increase of system
size, and approaches the algebraic form $J \sim 1/L$ in the thermodynamic
limit. For small $L$, however, $J$ varies slower than $1/L$.
Keeping this in mind, a suggestive phenomenological equation for the energy current can be 
written as
\begin{equation}
J = \kappa \frac{\Delta E}{L+\xi}
\label{xi}
\end{equation}
where $\kappa$ and $\xi$ are parameters which depend on $T, \Delta t$.
As the temperature $T \to 0$, the correlation length of the spin chain
$\xi \to \infty$, and consequently heat transport shows an apparent 
ballistic behavior.
In the other limit, i.e. when $T$ is large and $\xi \to 0$, thermal
transport in the system is diffusive.
 
\begin{figure}[htbp]
\centerline
{
\includegraphics[width=6.5cm,angle=-90]{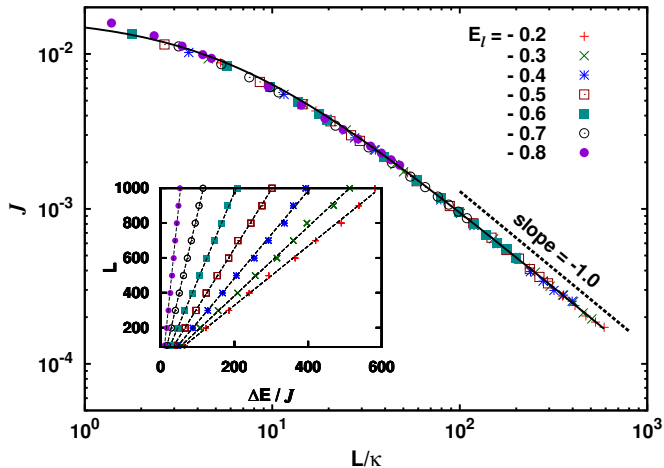}
}
\caption{(Color online) 
Collapse of the curves $J$ vs $L/\kappa$ for different values of average bath energies $E_l$ and $E_r$
with a fixed $\Delta E = 0.1$. The simulation data are shown by points and the solid line corresponds to the
curve of the form $\dfrac{\Delta E}{(L/\kappa) + \xi/\kappa}$. The $\kappa$ values are obtained from a straight
line fit of the form $L = \kappa \frac{\Delta E}{J} - \xi$ as shown in the inset (see text).
}
\label{fig:coll}
\end{figure}

Following Eq. (\ref{xi}), $\kappa$  and $\xi$ can be measured from the  
slope and intercept of the straight line  $L = \kappa \frac{\Delta E}{J} - \xi$.
In the inset of Fig. \ref{fig:coll}, we have shown $L$ against $\Delta E/J$ 
for different bath temperatures; all the curves are linear and the best fitted straight 
lines give respective $\kappa$  and $\xi$. Further, we observe that the parameters $\kappa$ and 
$\xi$ always maintain a fixed ratio with each other for any given $\Delta t$.
This becomes evident from the collapse of $J$ versus $L/\kappa$ curves
for different bath temperatures (see Fig. \ref{fig:coll}). This implies 
that $\kappa$ should have the same $T$ dependence as $\xi$.
Since near $T=0$ the correlation length $\xi \sim T^{-1}$, we expect that
$\kappa$ should also diverge inversely with $T$ in the limit $T \to 0$ .
In fact, $\kappa$ is the conductivity of the system in the thermodynamic limit $L\gg \xi$ 
and its divergence at $T=0$ indicates that the system is near a critical point.

\begin{figure}[htbp]
\centerline
{
\includegraphics[width=8.85cm,angle=0]{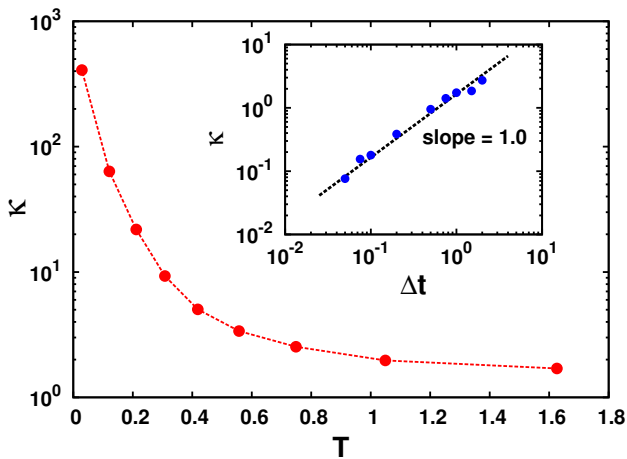}
}
\caption{(Color online) Variation of $\kappa$ with temperature $T$, as obtained 
from the straight line fit using Eq. (\ref{xi}), shows a divergence in 
$\kappa$ as $T \to 0$. (Inset) $\kappa$ varies linearly with $\Delta t$.
}
\label{fig:kappa}
\end{figure}

\begin{figure}[htbp]
\centerline
{
\includegraphics[width=6.4cm,angle=-90]{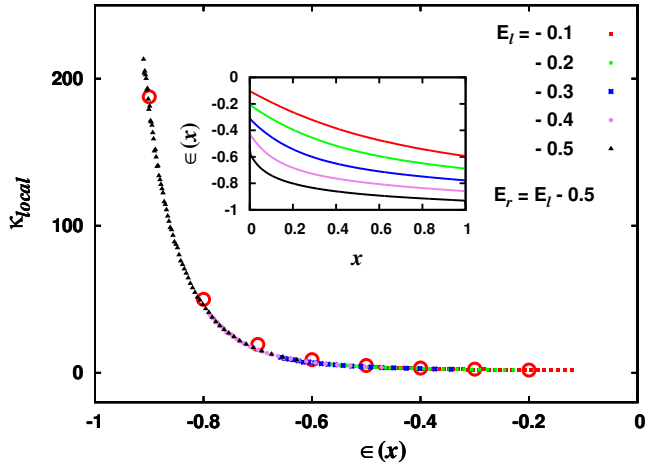}
}
\caption
{
Variation of local conductivity $\kappa_{local}$ with energy $\epsilon(x)$. $\kappa_{local}$ has 
been calculated locally (from the profiles shown in inset) using Eq. (\ref{localk}).
In the overlapping regions of the energy profiles, $\kappa_{local}$ 
for different energies collapses onto a single curve. As energy becomes small
(which corresponds to $T \to 0$), the local conductivity diverges. The open circles
correspond to $\kappa$ calculated using Eq. (\ref{xi}), $\Delta E = -0.1$, 
and for different $E_l = -0.2$ to $-0.9$ in steps of $-0.1$. For both the figures
$\Delta t = 1.0.$ and $L = 1000$.
}
\label{fig:localk}
\end{figure}

The behavior of $\kappa$ with temperature $T$ is shown in Fig \ref{fig:kappa}. Close to 
$T = 0$ the system relaxes extremely slowly and numerical studies in this limit 
become computationally expensive. One needs to go to extremely small temperatures
to see the $T^{-1}$ divergence of $\kappa$, which could not be reached with the 
available computational resources. The inset of Fig. \ref{fig:kappa} shows that 
$\kappa$ vanishes linearly in the limit $\Delta t \to 0$, which can be understood  
as follows. The DTOE update of a spin $\vec{S}_i$ keeps the local energy
$(\epsilon_{i-1} + \epsilon_i)/2$  conserved, i.e., $d\epsilon/dt = 0.$
Therefore, for finite $\Delta t$ we must have $\Delta \epsilon \sim (\Delta t)^2$
so that $\underset{\Delta t \to 0}{lim} \dfrac{\Delta \epsilon}{\Delta t} = 0$.
Again from Eq. (\ref{J}) we have
\begin{equation}
J \sim \langle \Delta \epsilon \rangle \sim (\Delta t)^2.
\label{JDeltat2}
\end{equation}
Since $\xi$ diverges as $(\Delta t)^{-1}$ (from Eq. (\ref{xiDt})) and the current in this limit
$J \sim  \frac \kappa \xi,$ we have $\kappa \sim \Delta t$.
Until now we have discussed thermal transport for a small $\Delta E$ and 
assigned the $\kappa$, obtained from Eq. (\ref{xi}), to be the conductivity of the
thermodynamic system at energy $E$ (or temperature $T$). As such, in this limit
the system is not {\it too} far from equilibrium, in a way that all parts of the system are 
maintained almost at the same temperature $T$. However if $\Delta E$ is appreciably larger, 
both local energy and its gradient varies significantly across the system.
In such a case, one can appropriately define a local conductivity as,
\begin{equation}
\kappa_{local} = J {\left( \dfrac{d\epsilon(x)}{dx}\right)_{local}^{-1}}.
\label{localk}
\end{equation}
To measure $\kappa_{local}$, we set the bath energies at $E_l$ and $E_r \ll E_l$ 
and calculate the energy profile $\epsilon (x)$  and its gradient 
$\dfrac{d\epsilon(x)}{dx}$ at different $x=i/L$ along the system. 
The inset of Fig. \ref{fig:localk} shows the energy profiles obtained for  
different bath energy $E_l$ and  $E_r= E_l - 0.5$.  In the main figure, we have shown
the local conductivity $\kappa_{local}$ as a function of the local energy $\epsilon(x)$;
the overlapping regions, although obtained from energy profiles with different boundary
energies, match remarkably.
Thus $\kappa_{local}$ is a well defined function of the energy (or equivalently, temperature)
and has the same value for a given energy, irrespective of the average energy of the two baths.
Since the spin system attains local thermal equilibrium for all nonzero temperatures,  
$\kappa_{local}$ at a given local energy $\epsilon$ must be same as the conductivity
$\kappa$ calculated using Eq. (\ref{xi}) for a large system with average bath energies 
$\epsilon$ and $\epsilon+\Delta E$ respectively. This is shown as open circles in 
Fig. \ref{fig:localk} for $\Delta E=0.1,$ and different $\epsilon.$

\subsection{Energy profiles}
We now turn to the energy profile of the driven system and investigate the dependence
of the same for the following three cases:
\\

\begin{figure}[htb]
\centerline
{
\includegraphics[width=7.75cm,angle=0]{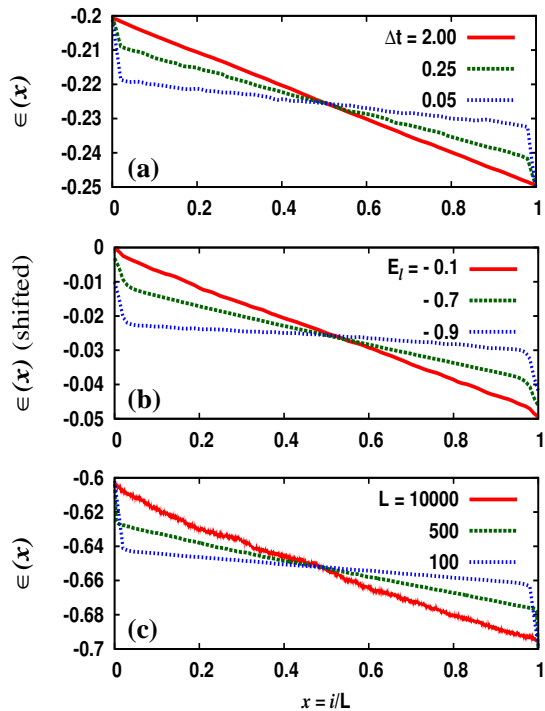}
}
\caption{(Color online) Energy profiles for 
(a) different $\Delta t$ with fixed bath energies $E_l = -0.2$, $E_r = -0.25$ and $L=100$.
(b) different bath energies $E_l,E_r$ with $\Delta E = -0.05$, $\Delta t = 0.1$ and $L = 100$. 
The profiles have been {shifted \textit{up} along the energy axis by $|E_l|$} to accommodate all the profiles 
within the same energy window.
(c) different $L$ with fixed bath energies $E_l = -0.6$, $\Delta E = -0.1$ and $\Delta t = 0.1$.
}
\label{fig:prof_ELdt}
\end{figure}

\paragraph{$\Delta t$ dependence.} 
The energy profile for a finite system depends
on the parameter $\Delta t$ as can be seen from Fig. \ref{fig:prof_ELdt}(a).
We have shown earlier (see section \ref{Drive} C) that for finite $L$, the two asymptotic limits $\Delta t \to 0, \infty$
correspond to ballistic and diffusive transport respectively and hence it is 
expected that for a smaller $\Delta t$ the profile will be relatively flatter
as compared to a larger value of $\Delta t$. Thus for any finite system if
$L \ll \xi$ the transport will be ballistic (flat energy profile) and one has
to simulate larger systems to observe a diffusive behavior (linear energy profile).

\paragraph{$E$ dependence.}
A lower $E$ implies a lower temperature $T$ and hence for a given $L$ and $\Delta t$, 
the correlation length monotonically increases as $E$ is decreased. The system
approaches a ballistic limit with energy profile as $E$ is reduced for a given
value of $\Delta t$ and $L$. However in the thermodynamic limit and for $T>0$,
Fourier law is always satisfied. This is shown in Fig. \ref{fig:prof_ELdt}(b)

\paragraph{$L$ dependence.}
The $L$ dependence of the energy profile is also consistent with what we have 
already discussed. For a given value of $\Delta t$ and $E$, a smaller $L$ shows
a flatter profile as compared to a system with larger $L$, as can be seen from
Fig. \ref{fig:prof_ELdt}(c).

\section{Discussion}
\label{Conclusion}
To summarize, we have studied thermal transport in a one-dimensional classical Heisenberg
spin model using discrete time parallel even-odd updates with spin precession (DTOE). 
While conventional integration schemes fail to preserve the required conservation of 
$S_i^2$ and $E$, this dynamics preserves both. 
{\clr The DTOE dynamics converts the equation of motion (\ref{eom})
to a {\it map} (Eq. (\ref{precess})) with an additional parameter $\Delta t$
(besides the interaction strength $K$ and the system size $L$); the equation of motion
is recovered from the map in the limit $\Delta t \to 0$.}
We explicitly show that this energy conserving dynamics 
equilibrates a closed system (having a fixed energy) and an open system attached
to equal temperature heat baths, for any finite $\Delta t$.
When the system is driven by maintaining a finite temperature difference between the two ends,
we explicitly show that the system attains local thermal equilibrium.
{\clr However, the steady state properties such as the correlation length $\xi$
(Eq. (\ref{xiDt})), thermal current (Eq. (\ref{JDeltat2})), and energy profile
(Fig. \ref{fig:prof_ELdt} (a)) depend on $\Delta t$ when the system size $L$ is finite;
such spurious $\Delta t$ dependence disappears in the thermodynamic limit.}

\begin{figure}[htb]
\centerline
{
\includegraphics[width=6.50cm,angle=-90]{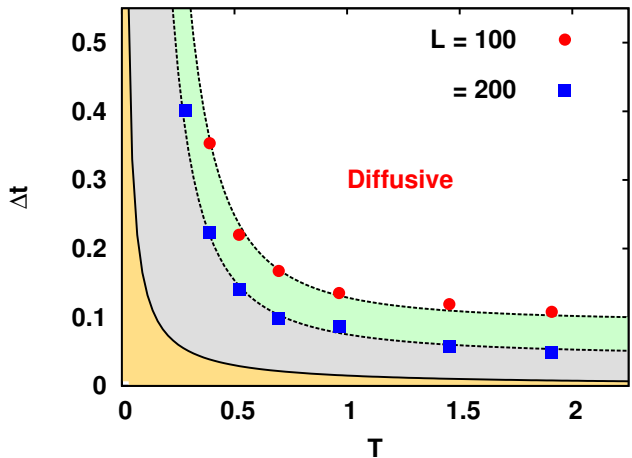}
}
\caption{{(Color online) The crossover line $T^*(\Delta t)$ in the $T$-$\Delta t$ plane for different $L$. 
In a finite system, this line separates the ballistic regime (shaded region below the curve), from the
diffusive one. The data points are obtained numerically (see text) for $L=100, 200$ (the broken line is only
a guide to the eye).
Following these trends, we schematically draw the same for very large $L$ (solid line). Evidently, the apparent 
ballistic behaviour disappears for $L\to \infty$ and one obtains Fourier law for all $T>0$, irrespective 
of the value of $\Delta t.$}
}
\label{fig:m}
\end{figure}

Our numerical simulations of the system for different bath temperatures
suggest that the thermal current $J$ can be expressed in the form 
$J = \kappa \frac{\Delta E}{L+\xi}$, where $\kappa$ and correlation length $\xi$
depend on the temperature $T$ and $\Delta t$.
{In the thermodynamic limit $L\gg \xi$, the spin system exhibits Fourier law; 
the energy profile in this case is linear with the  slope asymptotically 
approaching the value $m^*=\frac{\Delta E}{L}.$ However, for small system sizes 
(i.e. for $L \ll \xi$), thermal transport appears to be ballistic with a 
relatively flatter energy profile. The same scenario prevails when, instead, the 
temperature $T$ is varied. That is, for a  given $L$ and $\Delta t$, the slope of  
the energy profile approaches $m^*$ (or $0$) as $T\to \infty$ (or $0$).
Thus, finite systems show an apparent crossover from a diffusive to a ballistic 
behavior as $T$ is lowered below a characteristic temperature scale $T^*$.
This is described in Fig. \ref{fig:m} along with additional numerical evidences.    

To demonstrate the  crossover phenomena quantitatively, we measure the local slope
$m$ of the energy profile at $i=L/2$ for a system of size $L$ and fit it to a functional form 
\begin{equation}
m(T, \Delta t) = m^* \frac {T}{T + T^*(\Delta t)}.
\end{equation}
Clearly, $T^*$ is  the value of temperature for which the slope is half the desired  
slope for diffusive transport,  $m^*= \frac{\Delta E}{L}.$ 
In Fig. \ref{fig:m} we show the crossover temperature $T^*(\Delta t) $ for two 
different system sizes, $L=100, 200$. The crossover line  $T^*(\Delta t)$   
separates the  diffusive regime (well above the curve) from the  ballistic  
one  (well below the curve). Again, the crossover line shifts  
downwards when  the system  size $L$ is increased. This clearly indicates that 
in the thermodynamic limit, the  crossover  line  is infinitesimally close to 
the axes and thus, for any $T>0$, one observes a diffusive behaviour independent of 
the choice of $\Delta t.$
The apparent ballistic behaviour in the small $\Delta t$ 
or small $T$ regime is only an artifact of finiteness of the system 
and is a consequence of the divergence of the correlation length,
as $\xi \sim 1/T$ and $\xi \sim 1/\Delta t$.
The dependence of $\xi$ on $\Delta t$ can be understood from the fact that
the spin precesses by a small angle, proportional to $\Delta t$. This effect is 
similar to a low temperature precession-dynamics, where the correlation length 
$\xi$ is very large.
Thus, for studying thermal transport at a given $T$ and a small $\Delta t$,
one must carefully choose the system size to be large enough such that the
point ($\Delta t,T$) lies well above the crossover line.}

{The model is exactly solvable in both the limits $T\to 0$ and 
$T\to \infty$.}
In the $T\to 0$  limit, the thermal current $J = E_l- E_r$ 
is independent of $L$ and the energy profile $\epsilon(x)= \frac12(E_l+ E_r)$ 
is flat.  
In this limit Fourier law is violated as $J$ is not proportional to 
local slope of $\epsilon(x)$, which is now zero since $\epsilon(x)$ is flat.
The finite current $J = E_l - E_r$ is a consequence of the
{\it discontinuity } of the energy profile at the boundaries. 
This discontinuity, and therefore a finite current, can never be obtained
numerically for any finite $\Delta t$, however small. Numerical simulations,
in fact, show that the current vanishes as $J \sim \Delta t^2$ for small 
$\Delta t.$

In the other solvable limit, i.e. when $T\to \infty$, however, the energy profile
$\epsilon(x) = E_l + \frac{E_r-E_l}{L+2} x $ is linear and one obtains a
finite thermal conductivity $\kappa=1$.

In conclusion, a thermodynamically large classical Heisenberg spin chain in one dimension 
obeys Fourier law for any non-zero temperature.  
However while studying thermal transport numerically for a finite system (though large)
and a finite integration time step (though  small) one must be careful in keeping 
the boundary temperatures larger than the characteristic scale $T^*(\Delta t, L)$ to
obtain the correct thermodynamic behaviour. Otherwise, for $T \ll T^*$ the system will
show an apparent ballistic behaviour which will eventually disappear in the $L\to \infty$ limit.
This temperature dependent crossover from diffusive to ballistic behavior at small $T$ is expected 
since $T = 0$ is a critical point with a diverging correlation length.
It will be quite interesting to study thermal transport in a system where one can set the
boundary temperatures such that a singular point falls in the bulk.

\vskip0.25cm
{\it \bf Acknowledgement} : P.K.M. would thankfully acknowledge D. Dhar who
initiated this work and suggested this energy-conserving
dynamics, and M. Rao for fruitful discussions. The authors also wish
to thank the anonymous referees for providing constructive comments.

\end{document}